\documentclass[twocolumn,superscriptaddress,showpacs,preprintnumbers]{revtex4}
\usepackage{amssymb}
\usepackage{amsmath}
\usepackage[dvips]{graphicx}
\usepackage{dcolumn}
\usepackage{verbatim}
\usepackage{bm}

\begin{document}
\title{Division of the two-qubit Hilbert space according to the entanglement sudden death under composite noise environment}

\author{Peng Li}
\email{lipeng@fudan.edu.cn} \affiliation{Department of Physics and
Surface Physics Laboratory (National Key Laboratory), Fudan
University, Shanghai 200433, China} \affiliation{Department of
Materials Science, Fudan University, Shanghai 200433, China}

\author{Qun Zhang}
\affiliation{Department of Materials Science, Fudan University,
Shanghai 200433, China}

\author{J. Q. You}
\affiliation{Department of Physics and
Surface Physics Laboratory (National Key Laboratory), Fudan
University, Shanghai 200433, China}
\date{\today}

\begin{abstract}
We show theoretically that according to the disentanglement behavior
under composite noise environment, the Hilbert space of a two-qubit
system can be divided into two separate parts: a 3-dimensional
subspace in which all states disentangle asymptotically, and the
rest in which all states disentangle abruptly. The violation of
additivity for entanglement decay rates under weak noises [see, PRL
{\bf 97}, 140403 (2006)] therefore can be explained in terms of such
division of the Hilbert space.
\end{abstract}


\maketitle

\section{Introduction}\label{sec:1}
Quantum coherence, as a consequence of the superposition principle,
lies in the heart of quantum mechanics and is regarded as a unique
sign for the quantum regime. A fundamental question raised by
quantum coherence is the the emergence of macroscopic classicality
from the microscopic quantum world. It is currently well-recognized
that the answer lies in the so-called {\it decoherence} process, a
dynamical evolution of the system in which quantum coherence is
gradually lost due to the ubiquitous system-environment
interaction~\cite{decoherence}. Decoherence has been extensively
studied for years in the realm of quantum optics~\cite{Scully} as a
decay of {\it local} coherence. However, it was not until recently
that the evolution of {\it nonlocal} quantum coherence, i.e., the
{\it entanglement dynamics}, has become a field of interest that
attracts growing theoretical and experimental
studies~\cite{zyczkowski,Yu1,Yu2,Yu3,nonmarkovian,LFL,finitetem,Ficek,Goan,Lopez,Lastra,pairwise,ZSY,Konrad,Experiment}.
Moreover, since entanglement itself is a key resource for quantum
computation and information processing~\cite{Preskill,Nielson}, such
studies not only put new insights to the fundamentals of quantum
mechanics, but also scrutinize more stringently on whether and how
one can build an applicable quantum computer under realistic
circumstances.

A remarkable finding in the study of entanglement dynamics is the
entanglement sudden death (ESD), which was previously shown
in~\cite{zyczkowski} and latterly coined by Yu and
Eberly~\cite{Yu1,Yu2,Yu3}. The model studied in~\cite{Yu1} consists
of two initially entangled but non-interacting qubits each coupled
to its own local environment. In all circumstances the entanglement
between the two qubits disappears subsequently due to the
spontaneous decay of each qubit. Very interestingly, the
entanglement dynamics falls into two distinct categories: abrupt and
asymptotic disentanglement, depending on the initial states. The
abrupt disentanglement is characterized by a complete disappearance
of entanglement within finite time and is thus termed ESD. The
significance of ESD lies in two-folds: (i) It unveils a fundamental
difference between local and nonlocal quantum coherences via their
evolution dynamics; (ii) It puts an upper bound on the applicability
of entangled pairs in practical quantum communication even with the
best protocol for entanglement distillation.

Besides the apparent dependence on initial states, entanglement
dynamics is also intrinsically impacted by the specific form of the
system-environment interaction. Yu and Eberly also demonstrated
examples of ESD under composite noise environment~\cite{Yu2} and
classical noise environment~\cite{Yu3}. Structured reservoirs with
memory effects add a new ingredient, {\it entanglement revival}, to
the entanglement dynamics, due to the non-Markovian nature of the
system-reservoir interation~\cite{nonmarkovian}. Ikram {\it et al}
studied the effects of squeezed and thermal reservior~\cite{LFL}.
Al-Qasimi {\it et al}~\cite{finitetem} showed that all X-states
undergo ESD when the reservoir is at finite temperature. Extensions
of Yu and Eberly's model to commonly shared environment
(Dicke-regime)~\cite{Ficek} and high dimensional bipartite systems
(e.g., the $3\otimes3$ system~\cite{Lastra} and the continuous
variable system~\cite{Goan}) have also been studied recently.
Moreover, the question on where the lost entanglement goes has been
addressed recently~\cite{Lopez,pairwise}, revealing a rich and
counterintuitive relation between the ESD and the entanglement
sudden birth in the reservoirs.

However, in most
studies~\cite{Yu1,Yu2,Yu3,nonmarkovian,LFL,finitetem} the two-qubit
state under investigation is artificially confined to certain simple
classes (e.g., Bell-like states, Werner-like states, or the
X-states) for the convenience of theoretical handling. This led to a
lack of global knowledge on how the full 4-dimensional two-qubit
Hilbert space is structured according to the two distinct behaviors
of entanglement decay. From the perspective of quantum
communication, it is very important to know which state is more
robust than others in fighting against entanglement lost. Thus it is
desirable to discriminate states with asymptotic disentanglement
behavior from others. Very recently, Huang and Zhu~\cite{ZSY} have
specified the necessary and sufficient condition for ESD under
amplitude damping and phase damping, respectively, using a
principal-minors technique based on the positive partial transpose
(PPT) criterion. The purpose of this paper is to show that, even
under the combined action of these two noises, there exists a
3-dimensional ESD-free subspace, in which all state disentangles
asymptotically. Furthermore, we have proved that all states outside
this subspace disentangle abruptly under the composite noise
environment. Thus the full Hilbert space is completely divided into
two separate parts determined solely upon whether they undergo ESD
or not. Such a division of the Hilbert space also provides a simple
explanation to the violation of additivity for entanglement decay
rates under weak noises~\cite{Yu2} via the restriction of the
ESD-free subspace on the system dynamics.

This paper is organized as follows: In Section \ref{sec:2} we prove
that the combined action of amplitude damping and phase damping can
be factored as subsequent actions of each individually. Therefore
the set of ESD-free states under composite noise environment is
contained in the intersection of the two sets of ESD-free states
under each of the constituent noises. In Section \ref{sec:3} we show
that under pure phase damping, the ESD-free states constitute four
3-dimensional subspaces. In Section \ref{sec:4} we prove that only
one of the four subspaces is ESD-free under the composite noise
environment. Finally, we make a few remarks and give a short
conclusion in Section \ref{sec:5}.

\section{The factorization of amplitude damping and phase damping}\label{sec:2}
The system we study includes two non-interacting qubits (A and B)
and two independent local environments, which will be modeled as
either amplitude damping, or phase damping, or the combined action
of both.

The master-equation of the system under amplitude damping and phase
damping can be written as
\begin{equation}\label{me1}
\dot{\rho}=\sum_{A,B}\frac{\Gamma_1^{A,B}}{2}(2\sigma_-^{A,B}\rho\sigma_+^{A,B}
-\sigma_+^{A,B}\sigma_-^{A,B}\rho -
\rho\sigma_+^{A,B}\sigma_-^{A,B})
\end{equation}
and
\begin{equation}\label{me2}
\dot{\rho}=\sum_{A,B}\,\frac{\Gamma_2^{A,B}}{2}(\sigma_z^{A,B}\rho\sigma_z^{A,B}
-\rho),
\end{equation}
respectively. The two-qubit density operator $\rho$ is represented
in the computational basis $\vert 1\rangle=\vert
\uparrow\uparrow\rangle$, $\vert 2\rangle=\vert
\uparrow\downarrow\rangle$, $\vert 3\rangle=\vert
\downarrow\uparrow\rangle$, $\vert 4\rangle=\vert
\downarrow\downarrow\rangle$. The parameters $\Gamma_1^{A,B}$ and
$\Gamma_2^{A,B}$ (for party A, B) are the population relaxation rate
and dephasing rate for amplitude damping and phase damping,
respectively. The operators $\sigma_+^{A,B}$, $\sigma_-^{A,B}$ and
$\sigma_z^{A,B}$ denote the raising, lowering and Pauli operators
for each qubit. The explicit time evolution of Eq.(~\ref{me1}) and
(~\ref{me2}) can be solved and expressed in a unified form by the
Kraus-operators as
\begin{equation}
\rho(t)=\sum_{i=1,2,3,4}K_i \rho(0) K_i^\dag,
\end{equation}
with $\sum_i K_i^\dag K_i =1$. The two-qubit Kraus-operator $K_i$'s
can be expressed as tensor product of the one-qubit Kraus-operators
for each party. We specifically denote $K_i^{AM}$ as the two-qubit
Kraus-operator under amplitude damping defined by
$K_1^{Am}=M_1^A\otimes M_1^B$, $K_2^{Am}=M_1^A\otimes M_2^B$, $
K_3^{Am}=M_2^A\otimes M_1^B$, $K_4^{Am}=M_2^A\otimes M_2^B$, with
\begin{equation}
M_1^{A,B}=\left [ \begin{array}{cc} \gamma_1^{A,B} & 0\cr 0 &
1\end{array}\right], M_2^{A,B}=\left [ \begin{array}{cc} 0 & 0\cr
\omega_1^{A,B} & 0\end{array}\right],
\end{equation}
where $\gamma_1^{A,B}=\exp(-\frac{1}{2}\Gamma_1^{A,B}t)$ and
$\omega_1^{A,B}=\sqrt{1-(\gamma_1^{A,B})^2}$. While the
Kraus-operator $K_i^{Ph}$ for phase damping is defined by
$K_1^{Ph}=P_1^A\otimes P_1^B$, $K_2^{Ph}=P_1^A\otimes P_2^B$, $
K_3^{Ph}=P_2^A\otimes P_1^B$, $K_4^{Ph}=P_2^A\otimes P_2^B$, with
\begin{equation}
P_1^{A,B}=\left [ \begin{array}{cc} \gamma_2^{A,B} & 0\cr 0 &
1\end{array}\right], P_2^{A,B}=\left [ \begin{array}{cc}
\omega_2^{A,B} & 0\cr 0 & 0\end{array}\right],
\end{equation}
where $\gamma_2^{A,B}=\exp(-\Gamma_2^{A,B} t)$ and
$\omega_2^{A,B}=\sqrt{1-(\gamma_2^{A,B})^2}$. When both noises
participate, the master-equation reads
\begin{eqnarray}\label{me3}
\dot{\rho}&=&\sum_{A,B}\frac{\Gamma_1^{A,B}}{2}(2\sigma_-^{A,B}\rho\sigma_+^{A,B}
-\sigma_+^{A,B}\sigma_-^{A,B}\rho -
\rho\sigma_+^{A,B}\sigma_-^{A,B})\nonumber\\
&+&\sum_{A,B}\frac{\Gamma_2^{A,B}}{2}(\sigma_z^{A,B}\rho\sigma_z^{A,B}
-\rho).
\end{eqnarray}
Similarly, the time evolution of Eq.(~\ref{me3}) can be expressed by
the Kraus-operators as
\begin{equation}
\rho(t)=\sum_{k,l=1,2,3}\left(C_k^A\otimes C_l^B\right) \rho(0)
\left(C_k^A\otimes C_l^B\right)^\dag,
\end{equation}
where
\begin{equation}
C_1^{A,B}=\left [ \begin{array}{cc} \gamma_1^{A,B}\gamma_2^{A,B} &
0\cr 0 & 1\end{array}\right],\nonumber
\end{equation}
\begin{equation}
C_2^{A,B}=\left [ \begin{array}{cc} \gamma_1^{A,B}\omega_2^{A,B} &
0\cr 0 & 0\end{array}\right],\nonumber
\end{equation}
\begin{equation}
C_3^{A,B}=\left [
\begin{array}{cc} 0 & 0\cr \omega_1^{A,B} & 0
\end{array}\right].
\end{equation}

Using the above Kraus-operators it is straightforward to verify the
following equations
\begin{eqnarray}
&&\sum_{k,l=1,2,3}\left(C_k^A\otimes C_l^B\right) \rho(0)
\left(C_k^A\otimes C_l^B\right)^\dag\nonumber\\
&=&\sum_{j=1,2,3,4}K_j^{Am}\left[\sum_{i=1,2,3,4}
K_i^{Ph}\rho(0){K_i^{Ph}}^\dag\right]{K_j^{Am}}^\dag\nonumber\\
&=&\sum_{i=1,2,3,4}K_i^{Ph}\left[\sum_{j=1,2,3,4}
K_j^{Am}\rho(0){K_j^{Am}}^\dag\right]{K_i^{Ph}}^\dag,\nonumber\\
\end{eqnarray}
which can be further packaged into a compact form as
\begin{equation}\label{factor_channel}
\$^{C}_t=\$^{Am}_t\$^{Ph}_t=\$^{Ph}_t\$^{Am}_t,
\end{equation}
where $\$^{Am}_t$, $\$^{Ph}_t$, and $\$^C_t$ denote the quantum
channels governing the evolution of the system under amplitude
damping, phase damping, and both of them, respectively. The essence
of Eq.(~\ref{factor_channel}) is that the effect of the composite
noise environment can be factored as subsequent actions of each
constituent noise environment. Moreover, as a consequence of the
Markovian master-equations, it is also straightforward to verify a
general time-domain factorization property bellow
\begin{equation}~\label{factor_time}
\$_t=\$_{\tau'}\$_{\tau},
\end{equation}
where $t=\tau +\tau'$, $\tau,\tau'\geq0 $. Eq.(~\ref{factor_time})
holds for every kind of the quantum channels $\$^{Ph}_t$,
$\$^{Am}_t$ and $\$^C_t$. Since the entanglement is non-increasing
under local operations (note that all the quoted quantum channels
are local operations)~\cite{Preskill,Nielson}, a direct conclusion
drawn from the factorization properties Eq.(~\ref{factor_channel})
and (~\ref{factor_time}) is that if a state has completely
disentangled under either $\$_t^{Am}$ or $\$_t^{Ph}$, it must remain
disentangled under the action of $\$_t^{C}$. In other words, the set
of the ESD-free states under the composite noise environment is a
subset of the intersection of the sets of ESD-free states under each
one of the noises. Therefore, to be ESD-free under the action of
each noise is a necessary condition to be ESD-free under the action
of both noises.

Finally, we would like to clarify that amplitude damping itself does
not merely produce damping in the excited state of the qubit, but
also causes dephasing between the two states. This concomitant
dephasing has been automatically included in the the master-equation
(\ref{me1}) and the quantum channel $\$_t^{Am}$. In other words, it
is impossible to factor $\$_t^{Am}$ further into a fictitious ``pure
amplitude damping" channel and a fictitious ``pure dephasing"
channel. Therefore, when we talk about the composite noise
environment we mean that an extra dephasing noise, besides the
dephasing caused by amplitude damping, is present in the
system-reservoir interaction. Such composite noise environment is a
reasonable model for realistic circumstances where the noisy
environment can be eventually treated as a mixture of the two
noises.

\section{The ESD-free subspace under phase damping}\label{sec:3}
We have shown that to find the ESD-free states under the composite
noise environment, it is sufficient to explore within the set of the
ESD-free states under either one of the two noises. Our strategy is
to investigate the set of ESD-free states under phase damping first,
since there is a simple partition between the ESD and the ESD-free
states for pure phase noise. The question for the composite noise
environment will be addressed in the next section.

The necessary and sufficient condition for ESD under phase damping
has been specified by Huang and Zhu~\cite{ZSY} very recently using
the PPT criterion. The method they use is technically tedious and
lack of physical transparency. For the completeness of narration, we
re-derive it using a more intelligible method based on the
concurrence~\cite{Wooters}. The concurrence $C(\rho)$ is an
entanglement monotone ranges from $0$ to $1$ and can be computed
directly from the two-qubit density matrix $\rho$ as
\begin{equation}
C(\rho)=\max(0, \Lambda),
\end{equation}
where $\Lambda=\sqrt{\lambda_1}-\sqrt{\lambda_2}-
\sqrt{\lambda_3}-\sqrt{\lambda_4}$, with the $\lambda_i$'s the
eigenvalues of the matrix
\begin{equation}
R(\rho)=\rho\left(\sigma_y\
\otimes\sigma_y\right)\rho^\ast\left(\sigma_y\
\otimes\sigma_y\right)
\end{equation} in descending order.

We first consider the concurrence under phase damping in the
infinite-time limit, i.e., under the action of $\$^{Ph}_{\infty}$.
Suppose the initial state is
\begin{equation}
\rho(0)=\left[\begin{array}{cccc}
\rho_{11}   & \rho_{12} & \rho_{13} & \rho_{14} \\
\rho_{21}   & \rho_{22} & \rho_{23} & \rho_{24} \\
\rho_{31}   & \rho_{32} & \rho_{33} & \rho_{34} \\
\rho_{41}   & \rho_{42} & \rho_{43} & \rho_{44}
\end{array}\right].
\end{equation}
After infinite time of phase damping it is found
\begin{equation}
\rho(\infty)\,=\,\$^{Ph}_{\infty}\rho(0)\,=\,\left[\begin{array}{cccc}
\rho_{11}   & 0         & 0         & 0 \\
0           & \rho_{22} & 0         & 0 \\
0           & 0         & \rho_{33} & 0 \\
0           & 0         & 0         & \rho_{44}
\end{array}\right].
\end{equation}
The concurrence of the above final state is simply
\begin{equation}
C[\rho(\infty)]=\max\left[0, \Lambda(\infty)\right],
\end{equation}
where
\begin{equation}
\Lambda(\infty) = \left\{ \begin{array}{cc} -2
\sqrt{\rho_{11}\rho_{44}}, &
\rho_{22}\rho_{33} \geq \rho_{11}\rho_{44};\\
-2\sqrt{\rho_{22}\rho_{33}}, & \rho_{22}\rho_{33} <
\rho_{11}\rho_{44}.
\end{array}\right.
\end{equation}
Suppose $\rho_{11}\rho_{22}\rho_{33}\rho_{44}\neq0$, it is obvious
that
\begin{equation}\label{infinity}
\Lambda(\infty)<0.
\end{equation}
Since $\Lambda(t)$ is an algebraic function of the eigenvalues of
$R[\rho(t)]$, it must be a continuous function of $t$. The
inequality (\ref{infinity}) thus means that if $\Lambda(t)$ starts
from a positive value, it must have crossed over zero before
$t=\infty$, i.e., the two-qubit system must have disentangled at a
finite time. That is to say, the necessary condition for a state to
be ESD-free under phase damping is to have {\it at least one
vanishing diagonal element} in its density matrix. Thus any ESD-free
state must belong to one of the four subspaces below
\begin{displaymath}
\rho_{I}\,\,\,\,=\left[\begin{array}{cccc}
0 & 0         & 0         & 0         \\
0 & \rho_{22} & \rho_{23} & \rho_{24} \\
0 & \rho_{32} & \rho_{33} & \rho_{34} \\
0 & \rho_{42} & \rho_{43} & \rho_{44}
\end{array}\right],
\end{displaymath}
\begin{displaymath}
\rho_{II}\,\,\,=\left[\begin{array}{cccc}
\rho_{11} & 0 & \rho_{13} & \rho_{14} \\
0         & 0 & 0         & 0         \\
\rho_{31} & 0 & \rho_{33} & \rho_{34} \\
\rho_{41} & 0 & \rho_{43} & \rho_{44}
\end{array}\right],\nonumber
\end{displaymath}
\begin{displaymath}
\rho_{III}=\left[\begin{array}{cccc}
\rho_{11} & \rho_{12} & 0 & \rho_{14} \\
\rho_{21} & \rho_{22} & 0 & \rho_{24} \\
0         & 0         & 0 & 0         \\
\rho_{41} & \rho_{42} & 0 & \rho_{44}
\end{array}\right],
\end{displaymath}
\begin{equation}\label{subspaces}
\rho_{IV}=\left[\begin{array}{cccc}
\rho_{11}   & \rho_{12} & \rho_{13} & 0 \\
\rho_{21}   & \rho_{22} & \rho_{23} & 0 \\
\rho_{31}   & \rho_{32} & \rho_{33} & 0 \\
0           & 0         & 0         & 0
\end{array}\right].
\end{equation}

To prove that all states in $\rho_{I\sim IV}$ are indeed ESD-free
under phase damping, we further investigate the concurrence at
finite time $t$, e.g., for an initial state $\rho_I(0)$ located in
the subspace $\rho_I$. We find the eigenvalues of
$R[\rho_I(t)]=R[\$_t^{Ph}\rho_I(0)]$ are $\lambda_{1,2}=\alpha
\pm\beta$, $\lambda_{3,4}=0$, where
$\alpha=(\gamma_2^A\gamma_2^B)^2\rho_{23}\rho_{32}+\rho_{22}\rho_{33}$
and
$\beta=2\gamma_2^A\gamma_2^B\sqrt{\rho_{23}\rho_{32}\rho_{22}\rho_{33}}$.
Thus
\begin{eqnarray}\label{Lambda}
\Lambda(t)&=&\sqrt{\lambda_1}-\sqrt{\lambda_2}-\sqrt{\lambda_3}-\sqrt{\lambda_4}\nonumber\\
&=&2\beta/(\sqrt{\alpha+\beta}+\sqrt{\alpha-\beta}).
\end{eqnarray}
It is clear that $\Lambda(t)\geq0$ holds $\forall\, t$. Only in two
cases $\Lambda(t)=0$: (i) $t=\infty$, which means asymptotic
disentanglement; (ii) $\rho_{23}=\rho_{32}=0$, which means
$\Lambda(t)=0$ holds $\forall \, t$, i.e., the state $\rho_I(0)$ is
seperable from beginning. Thus all states in $\rho_{I}$ are indeed
ESD-free under phase damping. Straightforward calculations show that
the same conclusion holds for the other three subspaces $\rho_{II}$,
$\rho_{III}$ and $\rho_{IV}$. Thus, a generic state $\rho(0)$ is
ESD-free under phase damping iff
\begin{equation}
\rho_{11}\rho_{22}\rho_{33}\rho_{44}=0,
\end{equation}
or equivalent to say, iff it is located within the four subspaces
$\rho_{I\sim IV}$ shown in Eq.(~\ref{subspaces}).

\section{The ESD-free subspace under composite noise environment}\label{sec:4}
Now we further explore the entanglement dynamics for states in the
subspace $\rho_{I\sim IV}$ under the combined action of phase
damping and amplitude damping. We find completely different
disentanglement behaviors between the subspace $\rho_{I}$ and the
other three subspaces $\rho_{II\sim IV}$: all states in $\rho_I$
disentangle asymptotically and all states in $\rho_{II}$,
$\rho_{III}$ and $\rho_{IV}$ disentangle abruptly.

\subsection{All states in $\rho_{I}$ disentangle asymptotically}
It is straightforward to show that the subspace $\rho_I$ is full
ESD-free whatever the noisy environment is. For amplitude damping,
the evolution of an arbitrary initial state $\rho_I(0)$ is expressed
as $\rho_I'(t)=\$_t^{Am}\rho_I(0)$ and the eigenvalues of
$R[\rho_I'(t)]$ are $\lambda_{1,2}'=\alpha'\pm\beta'$ and
$\lambda_{3,4}'=0$, where
$\alpha'=(\gamma_1^A\gamma_1^B)^2(\rho_{23}\rho_{32}+
\rho_{22}\rho_{33})$ and
$\beta'=2(\gamma_1^A\gamma_1^B)^2\sqrt{\rho_{23}\rho_{32}\rho_{22}\rho_{33}}$.
Obviously, $\beta'\geq0$ and the equality holds only for $t=\infty$
or $\rho_{23}=\rho_{32}=0$. Following the same reasoning as for
Eq.(~\ref{Lambda}), it is clear that the subspace $\rho_{I}$ is also
ESD-free under amplitude damping. Similarly, when both noises are
present, the evoluation of $\rho_I(0)$ is simply
$\rho_I''(t)=\$_t^{C}\rho_I(0)$ and the eigenvalues of
$R[\rho_I''(t)]$ are $\lambda_{1,2}''=\alpha''\pm\beta''$ and
$\lambda_{3,4}''=0$, where
$\alpha''=(\gamma_1^A\gamma_1^B)^2[(\gamma_2^A\gamma_2^B)^2\rho_{23}\rho_{32}+
\rho_{22}\rho_{33}]$ and
$\beta''=2(\gamma_1^A\gamma_1^B)^2\gamma_2^A\gamma_2^B
\sqrt{\rho_{23}\rho_{32}\rho_{22}\rho_{33}}$. Again, $\beta''\geq0$
and the equality holds iff $t=\infty$ or $\rho_{23}=\rho_{32}=0$.
Thus $\rho_I$ is also ESD-free under composite noise environment.

\subsection{All states in $\rho_{II}$, $\rho_{III}$ and $\rho_{IV}$ disentangle abruptly}

The situations in $\rho_{II\sim IV}$ are more complicated because
the explicit expression of concurrence at finite time $t$ is too
lengthy to give a simple conclusion (due to the fact that the forms
of $\rho_{II\sim IV}$ do not preserve under amplitude damping). Such
complexity is also reflected in~\cite{ZSY}, where the derived
necessary and sufficient condition for ESD under amplitude damping
is mathematically packaged into the question of the positive
definiteness for an artificial $4\times 4$ matrix \{see Eq.(11)
of~\cite{ZSY}\}, which is still far from transparent.

However, using the factorization properties we can indeed prove that
no state in $\rho_{II\sim IV}$ is ESD-free under composite noise
environment. The trick is to divide the quantum channel $\$_t^{C}$
into two subsequent quantum channels $\$_\tau^C$ and $\$_{\tau'}^C$
as in Eq.(~\ref{factor_time}). The evolution of a generic intial
state $\rho(0)$ under $\$_t^C$ thus abides
\begin{equation}
\rho(t)=\$_t^C\rho(0)=\$_{\tau'}^C\$_\tau^C\rho(0)=\$_{\tau'}^C\rho(\tau),
\end{equation}
where $\rho(\tau)=\$_\tau^C\rho(0)$. Therefore the final state
$\rho(t)$ can be obtained by acting $\$_{\tau'}$ over the
intermediate state $\rho(\tau)$. Now one asks what is the necessary
and sufficient condition for asymptotic disentanglement for
$\rho(0)$ under the action of $\$_t^C$? The answer is quite simple:
{\it Neither has $\rho(\tau)$ completely disentangled within finite
time $\tau$, nor shall $\rho(\tau)$ disentangle completely under the
action of $\$_{\tau'}^C$, $\forall\, \tau,\tau'\geq0$}. The first
part of the statement is no more than a repeat of the original
question. However, the second part of the statement requires that
over the full time evolution any intermediate state $\rho(\tau)$
should belong to the set of the ESD-free states under the action of
$\$_{\tau'}^{C}$. Since the set of the ESD-free states under
$\$_{\tau'}^{C}$ is contained in the subspaces $\rho_{I\sim IV}$, it
is necessary for any intermediate state $\rho(\tau)$ to be located
within $\rho_{I\sim IV}$ to guarantee asymptotic disentanglement.
Now it is straightforward to show that as long as $\rho_{11}\neq0$,
such requirement cannot be satisfied whenever amplitude damping is
present for each party: the diagonal elements of $\rho(\tau)$ are
solely determined by amplitude damping as
\begin{eqnarray}
\rho_{11}(\tau)&=&(\gamma_1^A\gamma_1^B)^2\rho_{11},\nonumber\\
\rho_{22}(\tau)&=&(\gamma_1^A)^2[\rho_{22}+(\omega_1^B)^2\rho_{11}],\nonumber\\
\rho_{33}(\tau)&=&(\gamma_1^B)^2[\rho_{33}+(\omega_1^A)^2\rho_{11}],\nonumber\\
\rho_{44}(\tau)&=&\rho_{44}+(\omega_1^B)^2\rho_{33}
+(\omega_1^A)^2\rho_{22}+(\omega_1^A\omega_1^B)^2\rho_{11},\nonumber\\
\end{eqnarray}
where $\gamma_1^{A,B}$, $\omega_1^{A,B}$ are functions of $\tau$
instead of $t$. It is obvious that if $\rho_{11}\neq0$ and
$\Gamma_1^{A,B}\neq0$, none of the above diagonal elements vanishes
$\forall \, \tau>0$. Therefore, we arrive at the conclusion that no
state in $\rho_{II\sim IV}$ is ESD-free due to the existence of
nonzero element $\rho_{11}$ (Note that there exist states with
$\rho_{11}=0$ in $\rho_{II\sim IV}$, however, this trivial ambiguity
can be excluded by absorbing all these states into $\rho_{I}$).

\section{Discussion and conclusion}\label{sec:5}
An interesting topic connected with our work is the violation of
additivity for entanglement decay rates under weak noises discovered
by Yu and Eberly~\cite{Yu2}. They have demonstrated that a state
disentangles asymptotically under either amplitude damping or phase
damping may, however, disentangles abruptly under the combined
action of both. This fact unequivocally manifests the violation of
additivity for the decay rates of nonlocal quantum coherence, in
sharp contrast to the uphold of that for local quantum coherences.
In view of the ESD-free subspaces discussed, such phenomenon
deserves a simple explanation. The specific example used
in~\cite{Yu2} is no more than a special state located in the
subspace $\rho_{IV}$. Although it is ESD-free under either one of
the noises, it is not ESD-free under both because the evolution of
the state (due to amplitude damping) is not confined in the ESD-free
subspace required by the composite noise environment. According to
the previous analysis we conclude that such an abrupt violation of
additivity can only occur for states located in the subspaces
$\rho_{II\sim IV}$. On the other hand, due to the existence of the
completely ESD-free subspace $\rho_{I}$, it is still possible to
explore some quasi-additivity of entanglement decay rates for states
within this restricted space.

Another important work to note is the finite temperature effect
recently considered by Al-Qasimi and James~\cite{finitetem} for
amplitude damping (i.e., spontaneous excitation from
$\vert\downarrow\rangle$ to $\vert\uparrow\rangle$ is also allowed
due to thermal excitations in the reservoir). As demonstrated in an
X-state example, they argue that all states disentangle
asymptotically in zero-temperature bath must undergo sudden death at
finite temperature. This is a very strong statement that may smear
off the necessity to discriminate between the two kinds of
disentanglement behaviors under realistic circumstances. Therefore,
the significance of our work is restricted in the zero-temperature
regime, where spontaneous excitation is forbidden.

In conclusion, we have shown that the set of ESD-free states for a
two-qubit system under the combined action of amplitude damping and
phase damping is fully represented by the subspace $\rho_I$, which
is spanned on the bases $\vert\uparrow\downarrow\rangle$,
$\vert\downarrow\uparrow\rangle$ and
$\vert\downarrow\downarrow\rangle$. Therefore, any contamination
from the double-excitation state $\vert\uparrow\uparrow\rangle$ in
the initial density matrix is extremely hazardous for entanglement
preservation. Entanglement resources realized via
$\vert\uparrow\uparrow\rangle$ (e.g., the Bell-state $\phi_\pm$)
thus should be treated with great caution comparing to those without
it (e.g., the Bell-states $\psi_\pm$) whenever the lifetime of
entanglement has to be taken into consideration.

\acknowledgements This work was supported by the PCSIRT, the SRFDP,
the National Fundamental Research Program of China (NFRPC) Grant
No.~2006CB921205, and the National Natural Science Foundation of
China (NSFC) Grant No. 10625416 and No. 10534060.

\end{document}